\documentclass[twocolumn,superscriptaddress,bibnotes,showpacs,amsmath,amssymb,aps,prb]{revtex4-2}

\usepackage{graphicx}
\usepackage{amsmath}
\usepackage{amssymb}
\usepackage{color}
\usepackage{comment}
\usepackage[normalem]{ulem}
\usepackage{bm}

\usepackage[colorlinks=true,linkcolor=blue,citecolor=blue]{hyperref}

\begin{document}
\title{Bulk excitations in ultraclean $\alpha$-RuCl$_3$: Quantitative evidence for Majorana dispersions in a Kitaev quantum spin liquid}
\author{Kumpei~Imamura}\email{imamura@qpm.k.u-tokyo.ac.jp}
\author{Ryuichi~Namba}
\author{Riku~Ishioka}
\author{Kota~Ishihara}
\affiliation{Department of Advanced Materials Science, University of Tokyo, Kashiwa, Chiba 277-8561, Japan}
\author{Yuji~Matsuda}
\affiliation{Department of Physics, Kyoto University, Kyoto 606-8502, Japan}
\author{Seunghun~Lee}
\author{Eun-Gook~Moon}
\affiliation{Department of Physics, Korea Advanced Institute of Science and Technology (KAIST), Daejeon 34141, South Korea}
\author{Kenichiro~Hashimoto}
\author{Takasada~Shibauchi}\email{shibauchi@k.u-tokyo.ac.jp}
\affiliation{Department of Advanced Materials Science, University of Tokyo, Kashiwa, Chiba 277-8561, Japan}
\begin{abstract}

The spin-orbit coupled Mott insulator $\alpha$-RuCl$_3$ has emerged as a prime candidate for realizing the Kitaev quantum spin liquid (KQSL), characterized by Majorana quasiparticles, whose edge states exhibit a distinctive half-integer quantized thermal Hall conductivity. However, its van der Waals nature makes its thermal Hall response highly sensitive to structural disorder, leading to sample-dependent variations. Here, we investigate low-energy bulk excitations in the field-induced quantum disordered (FIQD) state of newly available ultraclean single crystals of $\alpha$-RuCl$_3$. High-resolution specific heat measurements under in-plane magnetic field rotation reveal an anisotropic excitation gap, whose field dependence is consistent with the Majorana gap in the KQSL state. Remarkably, when the field aligns with Ru-Ru bond directions, we observe gapless excitations with Dirac-like dispersions that quantitatively match theoretical predictions of Majorana bands based on the reported Kitaev interactions. Our findings in these ultraclean crystals provide strong evidence that the FIQD state of $\alpha$-RuCl$_3$ is a robust KQSL, resilient against small disorder perturbations.

\end{abstract}

\maketitle
The exactly solvable Kitaev model for insulating honeycomb magnets with bond-dependent Ising-type magnetic interactions realizes the Kitaev quantum spin liquid (KQSL) as a ground state~\cite{Kitaev2006}.
The excited states can be described by itinerant Majorana fermions and localized $Z_2$ fluxes (visons) due to the fractionalization of spins. In the presence of a magnetic field, bound states of these quasiparticles exhibit non-Abelian statistics, making them potential building blocks for topological quantum computation~\cite{Kitaev2006}.
The Kitaev interactions can be realized through the Jackeli-Khaliullin mechanism in real materials having edge-sharing octahedra surrounding magnetic ions~\cite{Jackeli2009}.
One of the prime candidates for realizing the KQSL is the spin-orbit assisted Mott insulator $\alpha$-RuCl$_3$~\cite{Takagi2019,Matsuda2025}.
The dominant interaction in $\alpha$-RuCl$_3$ is the ferromagnetic Kitaev exchange with a coupling strength of $J\sim5$-10\,meV~\cite{Suzuki2021,Maksimov2020}. However, the presence of additional non-Kitaev terms, including Heisenberg and off-diagonal exchange interactions, drives the system into an antiferromagnetic (AFM) ordered state below a N\'eel temperature of $T_{\text N}\sim7$ K~\cite{Johnson2015}.
Above $T_{\text N}$, signatures of the spin fractionalization consistent with the KQSL have been observed in various experiments such as neutron scattering~\cite{Banerjee2016proximate,Banerjee2017}, Raman scattering~\cite{Sandilands2015,Nasu2016}, and specific heat measurements~\cite{Do2017,widmann2019}. 
\par
\par
The AFM phase in $\alpha$-RuCl$_3$ is suppressed when the in-plane magnetic field exceeds $\sim 8$\,T, inducing the field-induced quantum disordered (FIQD) state~\cite{Yadav2016,Wolter2017,Banerjee2018}. Recent experimental investigations have reported the observation of half-integer quantized thermal Hall effect, a characteristic signature of chiral edge modes in the KQSL~\cite{Kasahara2018, Yokoi2021, Bruin2022, Yamashita2020,Xing2024}. However, definitive verification of the half-integer quantized thermal Hall effect remains a subject of rigorous investigation~\cite{Kasahara2022,czajka2023,lefranccois2022,Zhang2023_2,Zhang2024}. 
Complementarily, it has been discussed that specific heat measurements under in-plane field rotation can be used as a sensitive bulk probe of charge-neutral Majorana excitations with entropy because their low-energy dispersions have strongly field-angle dependent gap~\cite{Tanaka2022,Hwang2022,Imamura2024majorana,Fang2024}. In the FIQD state of $\alpha$-RuCl$_3$, an excitation gap is observed when the magnetic field is aligned along the zigzag directions ($\bm{H}\parallel\bm{a}$), which is closed along the Ru-Ru bond (armchair) directions ($\bm{H}\parallel\bm{b}$)~\cite{Tanaka2022,Imamura2024majorana}, demonstrating qualitative agreement with KQSL signatures. 
These findings, when combined with the in-plane field-angular dependence of thermal Hall conductivity, provide compelling evidence that Majorana fermion excitations are responsible for the observed planar thermal Hall effect. 
This observation substantiates the bulk-edge correspondence---a fundamental theoretical prediction of the topological KQSL state~\cite{Imamura2024majorana}.
\par
However, the van der Waals structure of $\alpha$-RuCl$_3$ makes it inherently susceptible to structural defects such as stacking faults~\cite{May_2020,Cao2016}. These structural defects, in conjunction with impurities, substantially influence various physical properties, including thermodynamic and thermal transport characteristics~\cite{Matsuda2025}. The manifestation of half-integer quantized thermal Hall effect exhibits pronounced sample-quality dependence~\cite{Kasahara2022,Zhang2023_2,Zhang2024}. Previous investigations predominantly utilized crystals synthesized via the chemical vapor transport (CVT) and Bridgman methods. 
Recently, a new two-step growth technique has been developed, which enables the synthesis of ultraclean single crystals with minimal structural defects and impurities~\cite{Namba2024}. Consequently, comprehensive thermodynamic measurements on these ultraclean crystals are essential for elucidating the intrinsic quasiparticle excitation properties of this system, which provides fundamental insights into the nature of the FIQD state.

\begin{figure}[b]
    \includegraphics[width=1\linewidth]{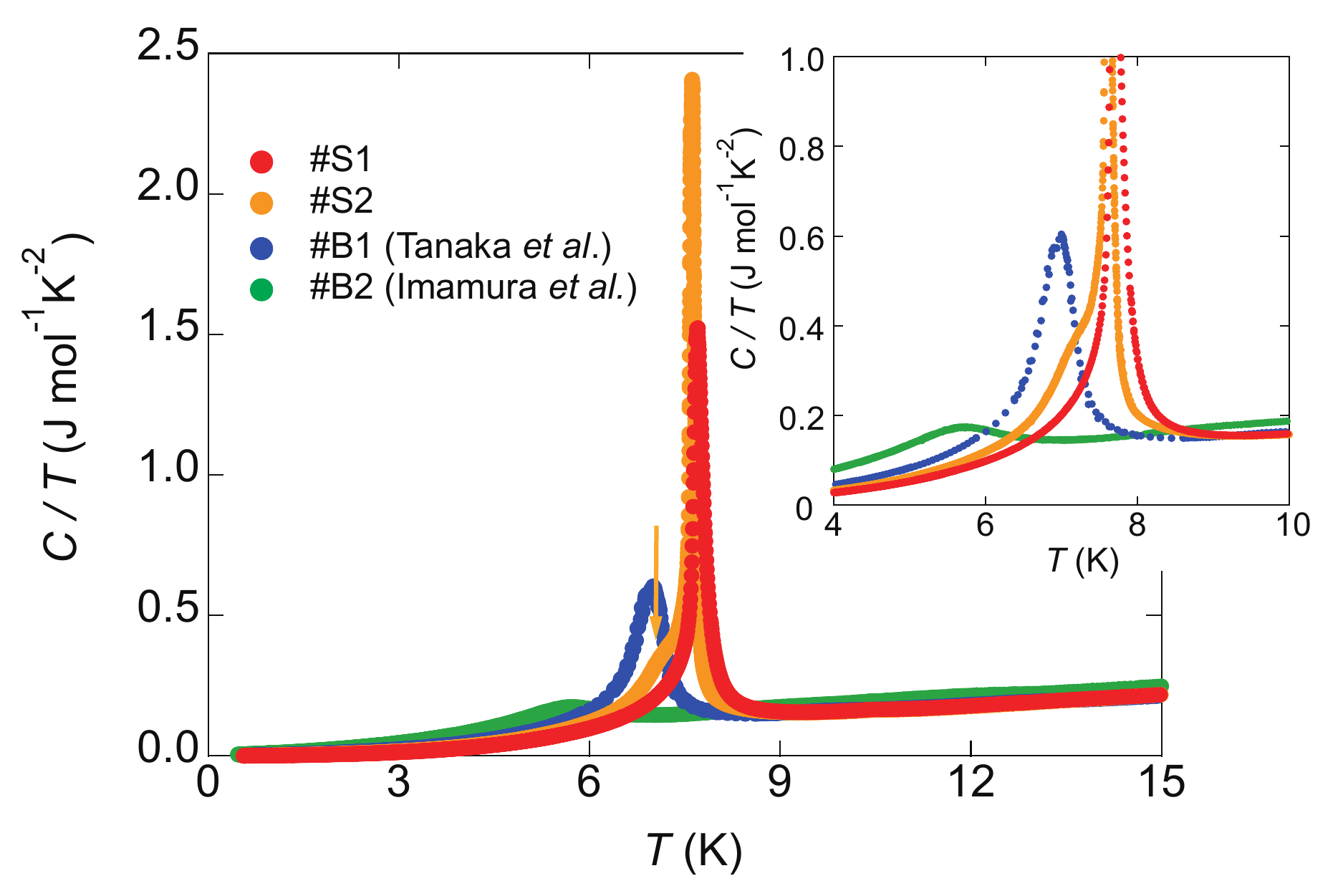}
    \caption{Temperature dependence of specific heat divided by temperature $C/T$ for the two-step sublimation samples \#S1 (red), \#S2 (orange) and Bridgman samples without (\#B1~\cite{Tanaka2022}, blue) and with (\#B2~\cite{Imamura2024defect}, green) electron irradiation. Orange arrow shows shoulder-like anomaly in \#S2. Inset: an enlarged view near the AFM transition.}
    \label{Fig1}
\end{figure}
\par

\begin{table*}[t]
    \centering
    \caption{Crystal dimensions (where the last number corresponds to the thickness perpendicular to the honeycomb plane), $T_{\mathrm{N}}$, $C/T$ at $T_{\mathrm{N}}$, and critical magnetic field $\mu_{\text 0}H_{\text c}$ for $\bm{H}\parallel\bm{a} (\bm{b})$ for $\alpha$-RuCl$_3$ samples.}
    \begin{tabular}{@{\hspace{1pt}}c@{\hspace{10pt}}c@{\hspace{10pt}}c@{\hspace{10pt}}c@{\hspace{10pt}}c@{\hspace{10pt}}c} \hline \hline
      Sample & dimensions ($\mu$m$^3$) & weight ($\mu$g) & $T_{\mathrm{N}}$ (K) & $C/T$ (J$\mathrm{mol^{-1}}$$\mathrm{K^{-2}}$) at $T_{\mathrm{N}}$ & $\mu_{\text 0}H_{\text c}$ (T) for $\bm{H}\parallel\bm{a} (\bm{b})$ \\
     \hline
       \#S1 & $1950\times1080\times77$ & 870 & 7.7 & 1.5 & 7.5 (8.0) \\
      \#S2 & $920\times1130\times120$ & 630 & 7.6 & 2.1 & 7.4 (8.0) \\
       \#B1~\cite{Tanaka2022} & $1210\times1380\times170$ & 680 & 7.0 & 0.6 & 6.9 (7.5) \\
       \#B2~\cite{Imamura2024defect} & $1110\times1280\times85$ & 340 & 5.7 & 0.2 & 6.5 (6.7)  \\ \hline \hline
    \end{tabular}
    \label{Table1}
 \end{table*}

Here, we investigate whether the anisotropic nature of Majorana excitations in the bulk state is indeed preserved in these ultraclean samples.
High-resolution measurements of the specific heat under in-plane field rotation reveal the presence of clear anisotropic bulk excitations in the ultraclean crystals. 
Additionally, by using samples with different $T_{\text N}$ values and mean free paths of heat carriers that correlate with each other, we examine the effect of sample quality on the bulk state. The anisotropy becomes smaller in cleaner samples, but the gapless excitations characterized by $C/T=\alpha T$ for $\bm{H}\parallel\bm{b}$ (parallel to $[\bar{1}10]$ in the spin coordinate) and the excitation gap sensitive to the field strength for $\bm{H}\parallel\bm{a}$ (parallel to $[11\bar{2}]$) are resolved even in the ultraclean samples. The quantitative comparisons with previous results reveal that, as the samples become cleaner, the coefficient $\alpha$ of the gapless state approaches the theoretically predicted value. Our analysis implies that the FIQD state of bulk $\alpha$-RuCl$_3$ in the clean limit is close to an ideal KQSL. 
\par
We use the ultraclean samples synthesized by the two-step method consisting of the pre-purification by CVT and the main growth by sublimation~\cite{Namba2024}. This CVT purification process can effectively remove impurities in the starting $\alpha$-RuCl$_3$ powder such as oxide phases, which is a key step in achieving ultraclean crystals. These samples exhibit a much sharper transition at higher temperatures $T_{\text N}$ compared to the Bridgman samples and a distinct first-order structural phase transition from the monoclinic ($C2/m$) to the rhombohedral ($R\overline3$) structure at $\sim 150$\,K. 
The specific heat $C(T,H)$ is measured in a $^3$He fridge by the long relaxation method, where a Cernox-1030 resistor acts as a thermometer, heater, and sample stage~\cite{Tanaka2022}. This experimental setup enables us to perform high-resolution specific heat measurements down to $\sim0.7$\,K under in-plane magnetic field rotation.

\begin{figure*}[t]
    \includegraphics[width=1\linewidth]{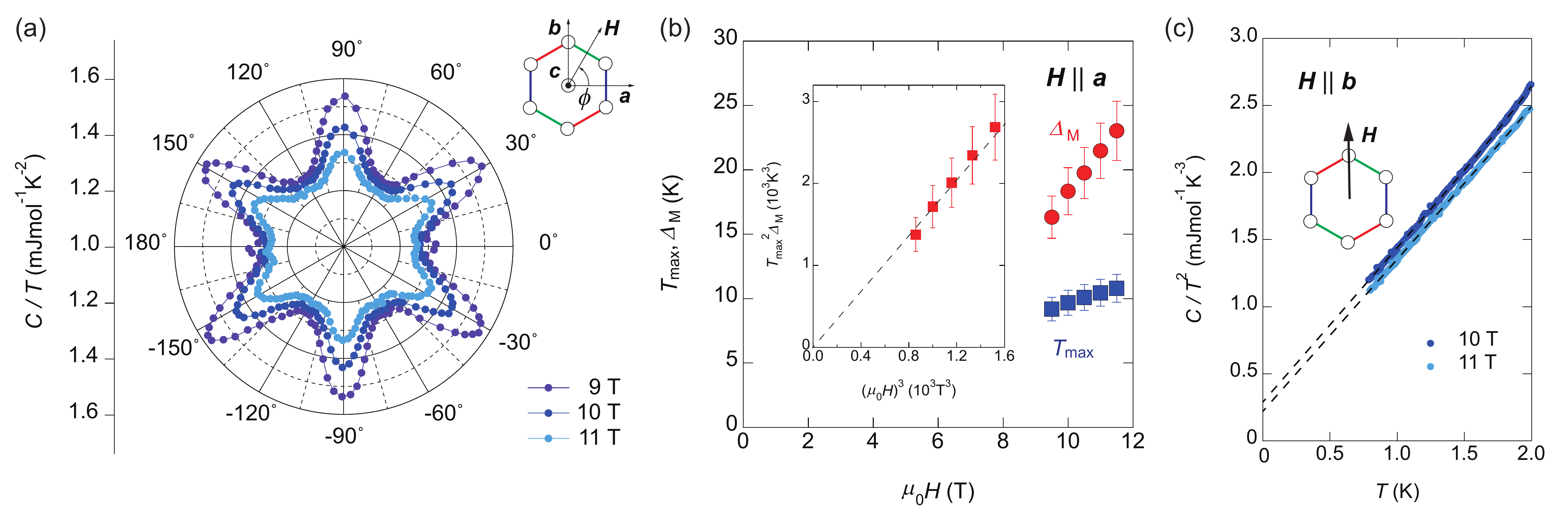}
    \caption{(a) In-plane field-angle dependence of $C/T$ at several fields at $T=1.0$\,K for the ultraclean sample \#S1. Inset defines the in-plane field angle $\phi$ from the $\bm{a}$ axis. (b) Field dependence of Majorana gap $\Delta_{\text M}$ (red circles) and $T_{\text{max}}$ (blue squares) for $\bm{H}\parallel\bm{a}$. The inset shows $T_{\text{max}}^2\Delta_{\text M}$ as a function of $H^3$. The dashed line represents the fitting curve for the expected $H^3$ dependence. (c) Temperature dependence of $C/T^2$ for $\bm{H}\parallel\bm{b}$ at 10 and 11\,T.}
    \label{Fig2}
\end{figure*}
\par
Figure\,1 shows the temperature dependence of specific heat divided by temperature $C/T$ for the two-step sublimation samples \#S1, \#S2 and Bridgman samples without (\#B1~\cite{Tanaka2022}) and with (\#B2~\cite{Imamura2024defect}) electron irradiation.
From the absence of anomalies around 10–14\,K in all samples, it is evident that the samples have very few stacking faults. On the other hand, the transition temperature $T_{\text N}$ and the specific heat jump at $T_{\text N}$ vary depending on the sample. The two-step sublimation samples exhibit higher $T_{\text N}$ and sharper, larger specific heat jumps at $T_{\text N}$ compared to the Bridgman samples. It is known that as sample quality deteriorates, $T_{\text N}$ decreases, and the specific heat jump is suppressed~\cite{Zhang2024}. Indeed, this behavior is observed in sample \#B2 where defects were artificially introduced through electron irradiation~\cite{Imamura2024defect}. From these observations, it is clear that the two-step sublimation samples have significantly higher quality than conventional samples such as the Bridgman samples. This is reinforced by the fact that the thermal conductivity in the AFM phase is much more enhanced than the Bridgman results, indicating that the mean free path of heat carriers is extremely long in these two-step sublimation crystals~\cite{Namba2024,Xing2024} (see also~\cite{SM}).
Table\,\ref{Table1} summarizes sample dimensions, $T_{\text N}$, $C/T$ at $T_{\text N}$, and the AFM critical magnetic field $H_{\text c}$, revealing that samples with higher $T_{\text N}$ tend to exhibit higher $H_{\text c}$. Sample \#S1 has the highest $T_{\text N}=7.7$\,K, indicating its superior quality. Sample \#S2 exhibits a similarly high $T_{\text N}\sim7.6$\,K without anomalies due to stacking faults and even higher $C/T$ at $T_{\text N}$, implying that the quality is close to that of \#S1. However, $C/T$ in \#S2 shows a shoulder-like feature around 7.1\,K below $T_{\text N}$ (see the inset of Fig.\,1), which suggests that sample \#S2 contains a small fraction part with slightly lower $T_{\text N}$. Thus, these results indicate that sample \#S2 is much cleaner than the Bridgman samples but has slightly more disorder compared to sample \#S1.
\par
First, we discuss the bulk low-energy excitations in the FIQD state of the ultraclean samples. Figure\,2(a) represents the in-plane field-angle dependence of $C/T(\phi)$ at $T=1.0$\,K at several fields in sample \#S1, where $\phi$ is the in-plane field angle from the $\bm{a}$ axis (inset of Fig.\,2(a)). $C/T(\phi)$ exhibits clear six-fold oscillations, taking the minimum for $\bm{H}\parallel\bm{a}$ ($\phi=0^\circ$) and the maximum for $\bm{H}\parallel\bm{b}$ ($\phi=90^\circ$). In contrast, in the magnetic phase, the opposite behavior is observed, with the maximum for $\bm{H}\parallel\bm{a}$~\cite{SM}. When the magnetic field is rotated within the honeycomb plane, the Majorana gap varies with a six-fold oscillation according to $\Delta_{\text M}\propto |\cos3\phi|$~\cite{Tanaka2022}. This can naturally explain the maximum $C/T$ for $\bm{H}\parallel\bm{b}$ (gapless direction) and the minimum $C/T$ for $\bm{H}\parallel\bm{a}$ (maximum gap direction). These observations are in qualitative agreement with the behavior seen in the Bridgman sample~\cite{Tanaka2022} and indicate that the six-fold oscillations in the FIQD phase match the anisotropy of the Majorana gap.
\par
To evaluate the Majorana gap for $\bm{H}\parallel\bm{a}$ where the Majorana gap is maximum, we perform the same fitting as used in the Bridgman sample previously~\cite{Tanaka2022} by three terms, $C(T,H)/T=\beta(H)T^2+C_{\text M}(T,H)/T+C_{\text{flux}}(T,H)/T$, representing the bosonic contribution, itinerant Majorana fermions, and $Z_2$ fluxes, respectively. 
Figure\,2(b) shows the field dependence of the Majorana gap $\Delta_{\text M}$ and $T_{\text{max}}$ obtained from the fitting. $T_{\text{max}}$ is the temperature at which $C-\beta T^3$ exhibits a peak~\cite{SM}, which is associated with the excitations of $Z_2$ fluxes~\cite{Do2017,Motome2020}. 
Considering the relationship $T_{\text{max}}\propto\Delta_{\text{flux}}$ found in quantum Monte Carlo simulations~\cite{Motome2020}, the Majorana gap and $T_{\text{max}}$ exhibit a characteristic field dependence described by $T_{\text{max}}^2\Delta_{\text M}\propto H^3$. We find that clear $H^3$ dependence of  $T_{\text{max}}^2\Delta_{\text M}$ holds in ultraclean samples (Fig.\,2(b), inset), which is well consistent with the results in the Bridgman sample~\cite{Tanaka2022}. 
\par
In the KQSL state, it is expected for $\bm{H}\parallel\bm{b}$ that the Majorana gap vanishes and the system shows a gapless linear Dirac dispersion ($E=v|{\bm k}|$), where $v$ represents the velocity in a two-dimensional system. 
In this case ($\bm{H}\parallel\bm{b}$), the low-energy excitations are represented by $C/T=\alpha T$, and indeed a finite $\alpha$ ($\lim_{T{\rightarrow}0}C/T^2 = \alpha$) has been observed in the Bridgman sample~\cite{Imamura2024majorana}. Figure\,2(c) represents the temperature dependence of $C/T^2$ in sample \#S1 without subtracting the phonon contribution. $C/T^2$ shows distinct residual values in the zero-temperature limit in the FIQD state above $H_{\text c}$. This indicates the presence of a finite $\alpha$ and gapless excitations even in the ultraclean sample. 

\par
Next, let us discuss the effect of the sample quality on the Majorana excitations in the bulk state more quantitatively. 
Figure\,3 represents the in-plane field-angle dependence of $C/T$ at $T=1.0$ K in samples \#S1 and \#S2. Sample \#S2 also shows similar sixfold oscillations consistent with the anisotropy of the Majorana gap, but the magnitude of the anisotropy is enhanced compared to \#S1. For $\bm{H}\parallel\bm{a}$, we find fully gapped excitations reflecting the Majorana gap in \#S2 as well. The obtained Majorana gap $\Delta_{\text M}$ is about 18.8\,K for 10\,T, which is almost the same value ($\sim18.3$\,K) in \#S1, and shows the characteristic $H^3$ dependence~\cite{SM}. This indicates that gapped excitations and the Majorana gap in the bulk of the KQSL are robust against small disorder. On the other hand, for $\bm{H}\parallel\bm{b}$, the magnitude of $C/T$ is changed noticeably. 
In theoretical calculations, it has been shown that introducing relatively strong disorder into the zero-field KQSL state results in a divergent low-energy density of states and induces local Majorana zero modes~\cite{Willans2010,Takahashi2023}. Furthermore, theoretical studies suggest that under a magnetic field, the disorder can lead to the emergence of gapless localized modes within the Majorana gap~\cite{Kao2021}. However, the effect of disorder on the gapless state for $\bm{H}\parallel\bm{b}$ has not been studied theoretically. The gapless state for $\bm{H}\parallel\bm{b}$ belongs to a different topological symmetry class compared to other directions \cite{Altland1997,O'Brien2016}, so it may exhibit different behavior under a magnetic field in response to disorder. In fact, in the Bridgman sample with electron irradiation, the disorder-induced low-energy excitations for $\bm{H}\parallel\bm{b}$ have been observed, which show stronger magnetic field dependence than the in-gap states for $\bm{H}\parallel\bm{a}$~\cite{Imamura2024defect}. 

\par

\begin{figure}[t]
    \includegraphics[width=1\linewidth]{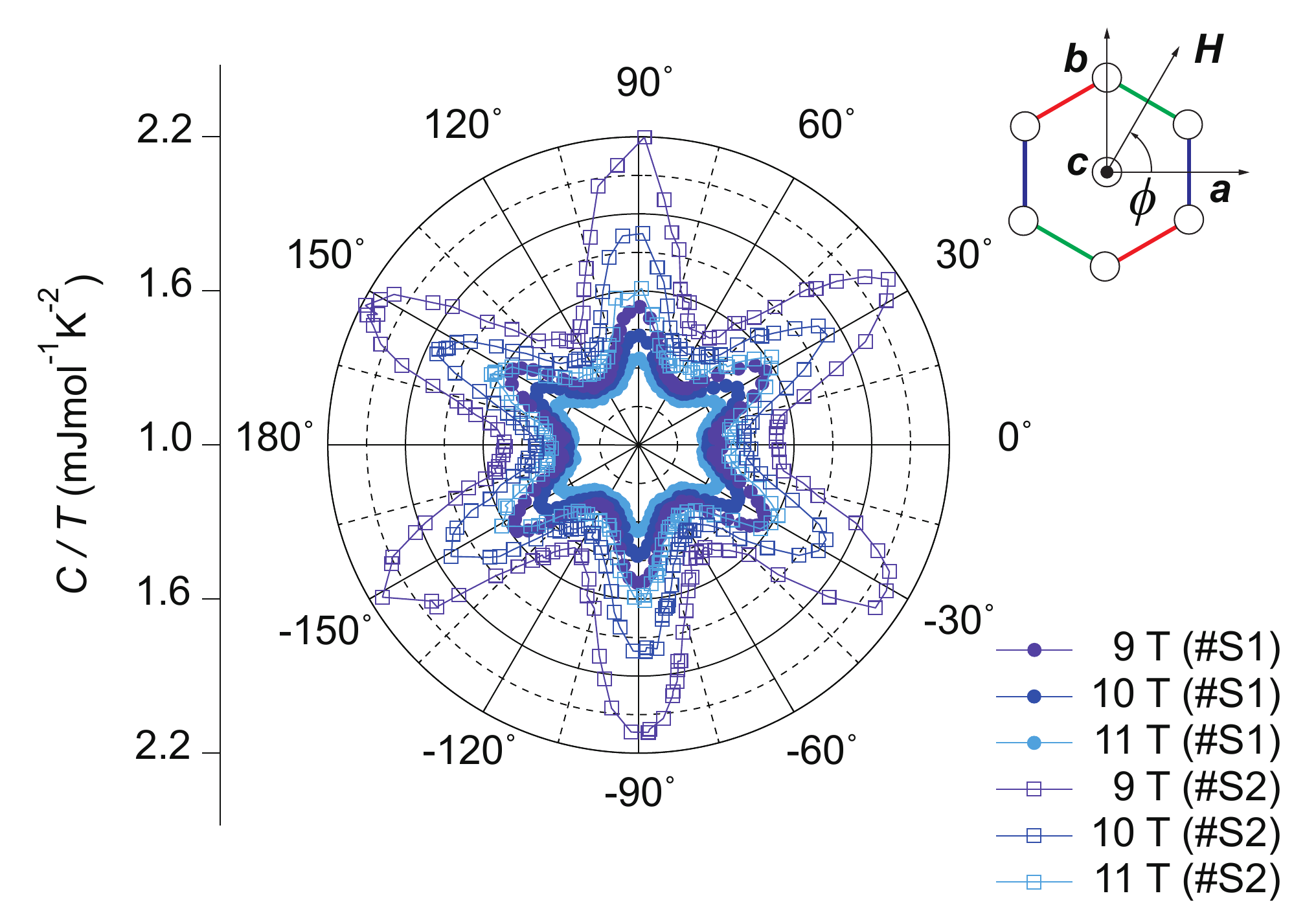}
    \caption{In-plane field-angle dependence of specific heat divided by temperature $C/T$ at 1\,K under several fields for samples \#S1 (solid circles) and \#S2 (open squares). }
    \label{Fig3}
\end{figure}

\par
We summarize the values of $\alpha$ at 10\,T ($\bm{H}\parallel\bm{b}$) for various samples with different $T_{\text N}$ values in Fig.\,4. As $T_{\text N}$ increases or the sample quality becomes higher, $\alpha$ decreases. This result implies that gapless excitations for $\bm{H}\parallel\bm{b}$ are sensitive to the cleanness of the crystals in the FIQD state of $\alpha$-RuCl$_3$. However, we emphasize that even in the ultraclean samples, $\alpha$ remains finite. The Kitaev interaction $J$, determined from several measurements and theoretical calculations, falls in a range between $\sim 5$ and $\sim 10$\,meV~\cite{Suzuki2021,Maksimov2020,Matsuda2025}. By using these values of  $J$, the Kitaev model leads to the estimate of $\alpha$ values ranging from $\sim0.15$ to 0.61\,mJmol$^{-1}$K$^{-3}$~\cite{SM}, as indicated by the gray hatch in Fig.\,4.
Therefore, the finite $\alpha$ of 0.3\,mJmol$^{-1}$K$^{-3}$ observed in the cleanest sample \#S1 is in quantitative agreement with the theoretically predicted value. 

\begin{figure}[t]
    \includegraphics[width=1\linewidth]{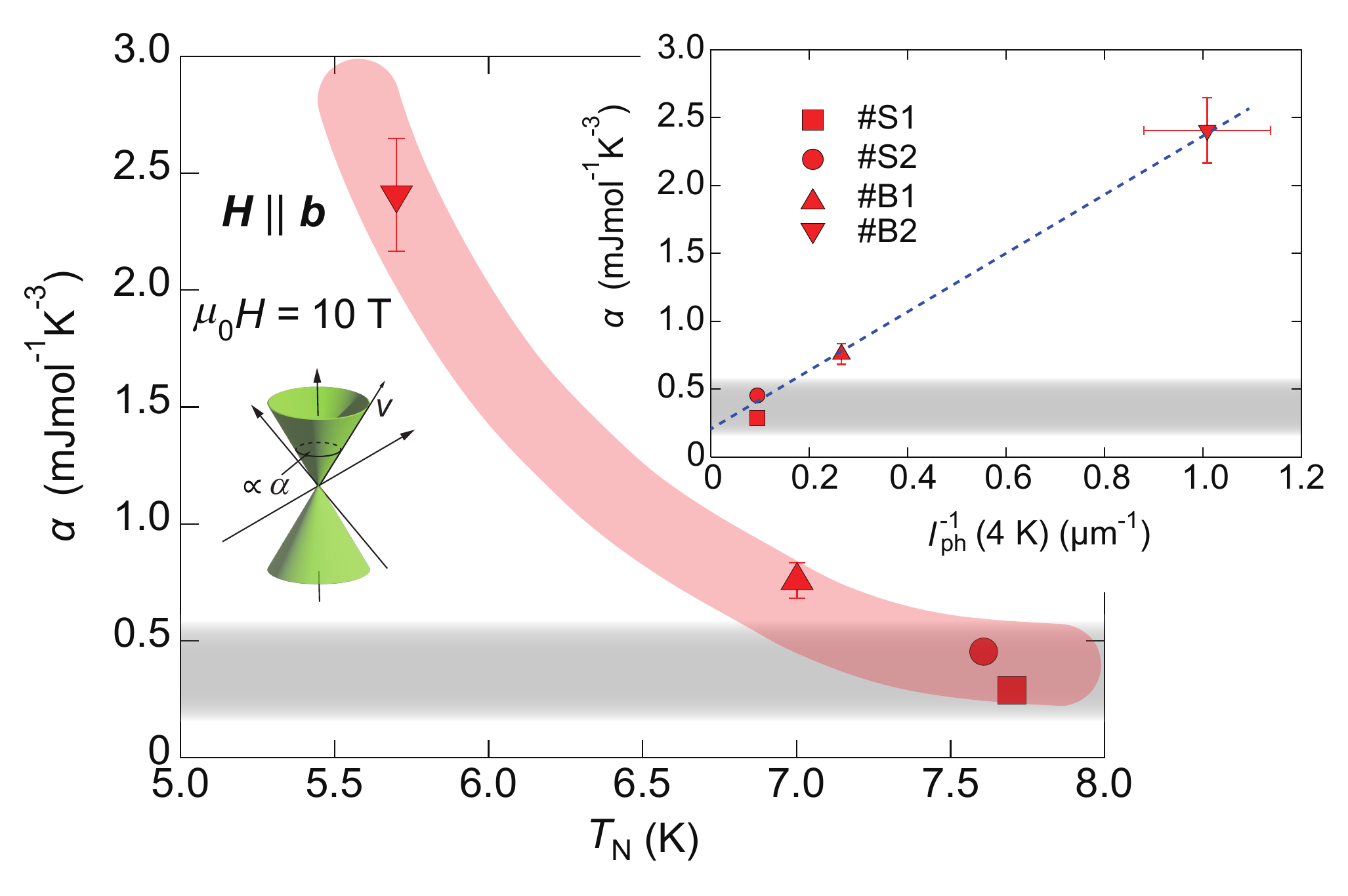}
    \caption{$T_{\text N}$ dependence of the coefficient $\alpha$ in the residual $T$-linear term of $C/T$ ($\alpha=\lim_{T{\rightarrow}0}C/T^2$) at 10\,T for $\bm{H}\parallel\bm{b}$. The gray hatch marks the theoretical estimate of $\alpha$, which is related to the Dirac-cone dispersion of Majorana fermions (inset sketch), using the reported Kitaev interaction $5\lesssim J \lesssim 10$\,meV. The inset shows $\alpha$ against the inverse of the phonon mean free path $l_{\text {ph}}$ at 4\,K~\cite{SM}. The dashed line is a guide to the eyes.}
    \label{Fig4}
\end{figure}

Furthermore, we compare our results with the reported sample dependence of the zero-field thermal conductivity $\kappa_{xx}$ in the AFM state, which is a good measure of cleanness. We use the  $\kappa_{xx}$ data for several samples, in which the stacking-fault anomalies are not pronounced, at 4\,K near the peak temperature of $\kappa_{xx}(T)$ where the phonon contribution dominates. From the simple relation $\kappa_{xx}=\beta_{\text {ph}}T^3v_{\text {ph}}l_{\text {ph}}/3$, where $\beta_{\text{ph}}$ is the coefficient of the phonon contribution in the specific heat and $v_{\text{ph}}$ is the sound velocity, we find a clear correlation between the phonon mean free path $l_{\text {ph}}$ and $T_{\text N}$~\cite{SM}. By using these estimates of $l_{\text {ph}}$ at 4\,K, we plot the gapless coefficient $\alpha$ as a function of the inverse of the phonon mean free path $l_{\text {ph}}$ (Fig.\,\ref{Fig4}, inset). This result reveals the close relation between the gapless excitations and the scattering rate. It also implies that the ultraclean samples with exceptionally long mean free paths are very close to the clean limit of $l_{\text{ph}}^{-1}\rightarrow0$. Indeed, the mean free path reaches $\sim10\,\mu$m, which is in the same order as the thickness of sample \#S1 (see Table\,\ref{Table1}). 
These results strongly suggest that the gapless excitations appearing only for $\bm{H}\parallel\bm{b}$, together with the robust field-sensitive gap for $\bm{H}\parallel\bm{a}$, are intrinsic properties in the FIQD phase of $\alpha$-RuCl$_3$. They reflect the characteristic field-angle-dependent excitations in the KQSL state.



Recently, thermal Hall conductivity measurements have been reported for ultraclean $\alpha$-RuCl$_3$ samples prepared by the two-step sublimation method~\cite{Xing2024}. Even in ultraclean samples, the thermal Hall conductivity exhibits a half-integer quantized plateau, consistent with the chiral Majorana modes in the edge state. Combined with the anisotropic bulk excitations revealed by the present specific heat measurements, this suggests that the bulk-edge correspondence characteristic of topological systems is realized even in the clean limit. On the other hand, the plateau region is narrower compared to that in Bridgman samples. This narrowing is attributed to the higher critical field $H_{\text c}$ and the increased phonon mean free path resulting from the superior sample quality, which facilitates decoupling between phonons and edge modes~\cite{Vinkler2018}. These findings underscore the importance of investigating not only the edge states, which are highly sensitive to sample disorder, but also the bulk excitations.
\par
In summary, we investigate the Majorana excitations in the bulk state of the KQSL using very high-quality $\alpha$-RuCl$_3$ samples. Our two-step method samples exhibit anisotropic behavior in $C/T$, reflecting the anisotropy of the Majorana gap. We observe a clear field-induced Majorana gap for $\bm{H}\parallel\bm{a}$ and low-energy gapless excitations characterized by $C/T=\alpha T$ for $\bm{H}\parallel\bm{b}$. Moreover, we investigate the effect of sample quality on the bulk state using samples with different $T_{\text N}$. As the samples become cleaner, the value of $\alpha$ for $\bm{H}\parallel\bm{b}$ decreases, and the anisotropy is suppressed. However, the gapless coefficient $\alpha$ for $\bm{H}\parallel\bm{b}$ remains finite even in the cleanest sample, showing a behavior that approaches the theoretically predicted value in the clean limit. These results provide quantitative evidence that the anisotropic Majorana excitations are an intrinsic property and that the KQSL state is realized in the FIQD phase of $\alpha$-RuCl$_3$. 

We thank Yuichi Kasahara and Yuta Mizukami for fruitful discussions. This work is supported by Grants-in-Aid for Scientific Research (KAKENHI) (No.\ JP23H00089) and for Transformative Research Areas (A) ``Correlation Design Science'' (No.\ JP25H01248) from the Japan Society for the Promotion of Science, and by CREST (No.\ JPMJCR19T5) from Japan Science and Technology Agency.

\par

\bibliography{Kitaev_ref}

\end{document}